\documentclass[3p]{elsarticle}

\usepackage{lineno,hyperref}
\modulolinenumbers[5]
\usepackage{textcomp}
\journal{Ocean Engineering}

\biboptions{authoryear}

\begin{document}

\begin{frontmatter}

\title{Comparing a 41-year model hindcast with decades of wave measurements from the Baltic Sea}


\author[FMI]{Jan-Victor Bj{\"o}rkqvist\corref{mycorrespondingauthor}}
\author[BMTARGOSS]{Ingvar Lukas}
\author[TUT]{Victor Alari}
\author[Delft]{Gerbrant Ph. van Vledder}
\author[BMTARGOSS]{Sander Hulst}
\author[FMI]{Heidi Pettersson}
\author[HZG]{Arno Behrens}
\author[TUT,EEA,UT]{Aarne M\"annik}
\cortext[mycorrespondingauthor]{Corresponding author, email: jan-victor.bjorkqvist@fmi.fi.\\ Final article: \url{ https://doi.org/10.1016/j.oceaneng.2018.01.048}}

\address[FMI]{Finnish Meteorological Institute, Marine Research, P.O. Box 503, FI-00101 Helsinki, Finland}
\address[BMTARGOSS]{BMT ARGOSS, Spacelab 45, Amersfoort, 3824 MR, The Netherlands}
\address[TUT]{Tallinn University of Technology, Department of Marine Systems, Akadeemia tee 15a, 12611, Tallinn, Estonia}
\address[Delft]{Delft University of Technology, P.O. Box 2048, 2600 GA, Delft, The Netherlands}
\address[HZG]{Helmholtz-Zentrum Geesthacht, Max-Planck Strasse 1, 21502 Geesthacht, Germany}
\address[EEA]{Estonian Environment Agency, Mustam\"ae tee 33, 10616 Tallinn, Estonia}
\address[UT]{University of Tartu, \"Ulikooli 18, 50090 Tartu, Estonia}

\begin{abstract}
We present ice-free and ice-included statistics for the Baltic Sea using a wave hindcast validated against data from 13 wave measurement sites. In the hindcast 84\% of wave events with a significant wave height over 7 m occurred between November and January. The effect of the ice cover is largest in the Bay of Bothnia, where the mean significant wave height is reduced by 30\% when the ice time is included in the statistics. The difference between these two statistics are less than 0.05 m below a latitude of 59.5\textdegree. The seasonal ice cover also causes measurement gaps by forcing an early recovery of the instruments. Including the time not captured by the wave buoy can affect the estimates for the significant wave height by roughly 20\%. The impact below the 99\textsuperscript{th} percentiles are still under 5\%. The significant wave height is modelled accurately even close to the shore, but the highest peak periods are underestimated in a narrow bay. Sensitivity test show that this underestimation is most likely caused by an excessive refraction towards the shore. Reconsidering the role of the spatial resolution and the physical processes affecting the low-frequency waves is suggested as a possible solution.
\end{abstract}

\end{frontmatter}


\section{Introduction}
Knowledge of the sea state is essential for diverse engineering, oceanographic and climatological purposes. Model simulations covering multiple spatial and temporal scales are a common method for acquiring the spatio-temporal characteristics of wave parameters. The well known KNMI/ERA-40 global wave atlas \citep{Sterl_Caires_2005} satisfactorily describes the wave climate of the World Ocean and has also been used to calculate exceedance values for significant wave height. It is, however, not intended to resolve regional wave climates, such as the Baltic Sea climate.

The Baltic Sea is a semi-enclosed body of water ranging from 9\textdegree --30\textdegree\ E to 53\textdegree--66\textdegree\ N and it is characterised by a seasonal ice cover. It has several topographically and geographically defined sub-basins with a combined area of 435,000 $\textrm{km}^2$ and a longest possible fetch of about 700 km (Fig. \ref{Fig:iow_bal}). While the mean water depth is only 55 metres the maximum depth reaches 459 metres. The Baltic Sea has heavy marine traffic \citep{HELCOM2010}, but wave data is also in demand for coastal planning purposes \citep[e.g.][]{Kahmaetal2016}.

The wave climate of the Baltic Sea has been assessed using both instrumental wave measurements \citep[e.g.][]{Kahma2003, PetterssonHELCOM2004, Broman2006} and model hindcasts \citep[e.g.][]{Jonsson2003, Raamet2010,Tuomi2011}. Both approaches have their limitations. The measurements can be lacking in terms of spatial coverage, especially since almost no instrumental wave measurements exist from the central and eastern Baltic Proper. There are also no wave measurements from the Gulf of Riga, except for short measurement campaigns \citep{Suursaar_etal_2012}. 
In the eastern Baltic Sea region there are no continuous instrumental wave measurements either, but only visual estimates made by observers onshore \citep{Soomere2013}. These observations, however, do not represent open sea conditions and are lacking homogeneity in time. Only one long instrumental time series spanning several decades can be found for the southern Baltic Sea region \citep{Soomere_etal_2012}. The majority of instrumental observations in the Baltic Sea are made with wave buoys. Because of the seasonal ice-cover wave buoy measurements seldom cover the entire ice-free period, since the devices have to be removed in advance to avoid damage by freezing. This adds one more factor to take into account when considering the representability of measurements. 

Model hindcasts are able to provide spatial information about the wave field, but the resolutions used in previous studies ($\sim$6--11 km) might not replicate all its features with sufficient accuracy. Near shore conditions in particular are still a big challenge for wave models \citep{Tuomi2014,Bjorkqvist2017b}. Not all hindcast studies include the ice-cover \citep{Jonsson2003, Raamet2010}, while other studies have even used daily updated ice-charts \citep{Tuomi2011}. The quality of the wind forcing is also a limiting factor, and the resolution used in different studies has varied from 9 km \citep{Tuomi2011} to 111 km \citep{Raamet2010}. The hindcast lengths for the whole Baltic Sea ranges from 1 year \citep{Jonsson2003} to 43 years \citep{Cieslikiewicz2008}. Recently, \citet{Siewert2015} hindcast the western Baltic Sea wave fields with a 52-year simulation. 

The aim of this paper is to use a new high-resolution ($\sim$1.85 km) simulation to present more accurate long-term (41 years) wave statistics for the Baltic Sea. Together with several extensive observational data sets from three different institutes, we are also able to study the similarities and differences between the wave statistics when estimating return values based on measured and modelled time series of different lengths. We will focus especially on the limitations a seasonally ice-covered sea impose on the measurements by quantifying the impact of the resulting measurement gaps. \citet{Caires2005} limited the ERA-40 data set so that it would always match the wave buoy measurements, thus not quantifying the statistics lost by the gaps. They also completely excluded years with gaps longer than one month. This approach is too strict in the Baltic Sea area, which has an ice cover that can last several months per year. 

The paper is structured as follows. In Section \ref{sec:MM} we introduce the wave model set-up, the atmospheric forcing and the wind and wave measurements used in this study. Section \ref{sec:Validation} presents an extensive validation of the wave model results covering all the different sub-basins of the Baltic Sea, except for the Gulf of Riga. The wave statistics from the model hindcast are presented in Section \ref{sec:Wave_stat}, while the difference in determining wave height and wave period exceedance values from both the measurements and the hindcast is explored in Section \ref{sec:Comp}. Conclusions are formulated in Section \ref{sec:Conclusions}.

\section{Materials and methods}\label{sec:MM}

\subsection{The wave model SWAN}
We used the wave model SWAN cycle III (version 41.10) to generate the Baltic Sea wave hindcast for 1965--2005. The wave model SWAN \citep{Booij1999} is a third-generation phase-averaged spectral wave model that was developed at Delft University of Technology. The waves are described via the two-dimensional wave action density spectrum, $N$, the evolution of which is governed by the wave action balance equation. This equation, in Cartesian coordinates without ambient currents, takes the following form:

\begin{center}
\begin{equation} \frac{\partial N}{\partial t} + \frac{\partial c_x N}{\partial x}+\frac{\partial c_y N}{\partial y}+\frac{\partial c_{\sigma} N}{\partial \sigma} + \frac{\partial c_{\theta} N}{\partial \theta} = \frac{S_{tot}}{\sigma}. \label{Eq:EBE}\end{equation}
\end{center}

The terms on the left represent the rate of change and the propagation of wave energy in two-dimensional geographical space, as well as the shifting of the radian frequency caused by variations in depth and depth-induced refraction. The x- and y-components of the group velocity are denoted by $c_x$ and $c_y$. The propagation velocities in the spectral space, which is defined by the radian frequency ($\sigma$) and the propagation directions ($\theta$), are $c_{\sigma}$ and $c_{\theta}$ respectively. Expressions for the spectral velocities can be found in the SWAN technical manual \citep{SWANtech}.

The term on the right-hand side, $S_{tot}$, contains the source terms that represents all physical processes that generate, dissipate or redistribute wave energy in SWAN. It is divided into six different terms. The deep water source terms are the energy input by wind \citep{Komen1984}, the dissipation of waves by whitecapping \citep{Komen1984}, and the nonlinear transfer of wave energy due to four-wave interactions using the Discrete Interaction Approximation \citep[DIA,][]{Hasselmann1985}. The whitecapping coefficient $\delta$ was set at 1 following \citet{Rogers2003} and \citet{Pallares2014}. We used the wind drag parametrisation suggested by \citet{Wu1982}, since we found a strong negative bias when using the default drag in SWAN proposed by \citet{Zijlema2012}. The shallow-water source terms are the energy dissipation through bottom friction \citep{Hasselmann1973}, dissipation due to depth-induced wave breaking \citep{BJ78} and the nonlinear transfer of wave energy through three-wave interactions using the Lumped Triad Approximation \citep[LTA,][]{Eldeberky1996}. The bottom friction coefficient was set at 0.038 m$^2$s$^{-3}$, as suggested by \citet{Zijlema2012}, and the parameter values of $\alpha$ and $\gamma$ for the depth-induced wave breaking source term were set at 1 and 0.73, respectively.

The model was run with a 15-min integration time step and using one iteration per time step. The structured grid had spherical coordinates with a resolution of 1' latitude and 2' longitude, and it extended from 9.016\textdegree\ E to 30.983\textdegree\ E and from 53.508\textdegree\ N to 65.99\textdegree\ N, yielding 660 x 750 grid points. The wave spectrum in SWAN consisted of 36 equally spaced directions and 32 frequencies distributed logarithmically on the frequency range 0.05--1 Hz. In this study, we generated output for the significant wave height ($H_s$) and the peak period ($T_p$), but the hindcast also contains, e.g. mean periods and directional parameters, all at a 1 h interval. In that respect, this database represents an advancement over the one generated by \citet{Suursaar2014}. The significant wave height, $H_s$, is defined as 

\begin{center}
\begin{equation}H_s=H_{m_0}=4\sqrt{m_0}, \label{Eq:Hm0}\end{equation}
\end{center}

where $m_0$ is the zero-order moment of the one-dimensional wave spectrum. The peak period ($T_p$) is defined as the period of the wave spectrum containing the most energy. 

To quantify the differences between the measured and modelled parameters, we used the established parameters bias, root-mean-square error (RMSE) and Pearson correlation coefficient. A positive bias in this paper means that the wave parameters are overestimated in the hindcast.

\subsection{The forcing fields}
The wind fields used in this study originate from the Baltic Sea regional reanalysis database BaltAn65+ \citep{Luhamaa2011}, which is a regional refinement of the ERA-40 and ERA Interim data sets that covers the period 1 January 1965 to 31 December 2005. The atmospheric model HIRLAM \citep{HIRLAM} was used for the reanalysis. Wind data, which was available every 6\textsuperscript{th} hour, were interpolated in their components internally according to the integration time step in SWAN. The horizontal grid resolution of wind forcing was 0.1 degrees (approximately 11 km).

A coupled ice-ocean model from the Swedish Meteorological and Hydrological Institute (SMHI) supplied the ice concentrations for the wave model. The two coupled models were the Rossby Centre Ocean (RCO) model and the Helsinki Multicategory Sea Ice Model (HELMI) \citep{Haapala2005}, which ran for the entire hindcast period with a resolution of two nautical miles ($\sim$3.7 km). The coupled ice-ocean model was forced with the downscaled ERA-40 data set, which have a 25 km resolution. A validation of the results is presented in \citet{Loptien2013}. The ice model recorded the ice concentration every two days. Grid points exceeding a concentration of 50\% were assigned as dry points in the wave model. We realize that this way of treating the ice effect is rather crude, but more advanced approaches are still not properly validated.

We used a digital topography covering the entire Baltic Sea with a resolution of 1 nautical mile \citep{Seifert2001}. Currents and spatial varying water levels were not included in the model used for this study.

\begin{figure}
\begin{center}
\includegraphics[width=14cm]{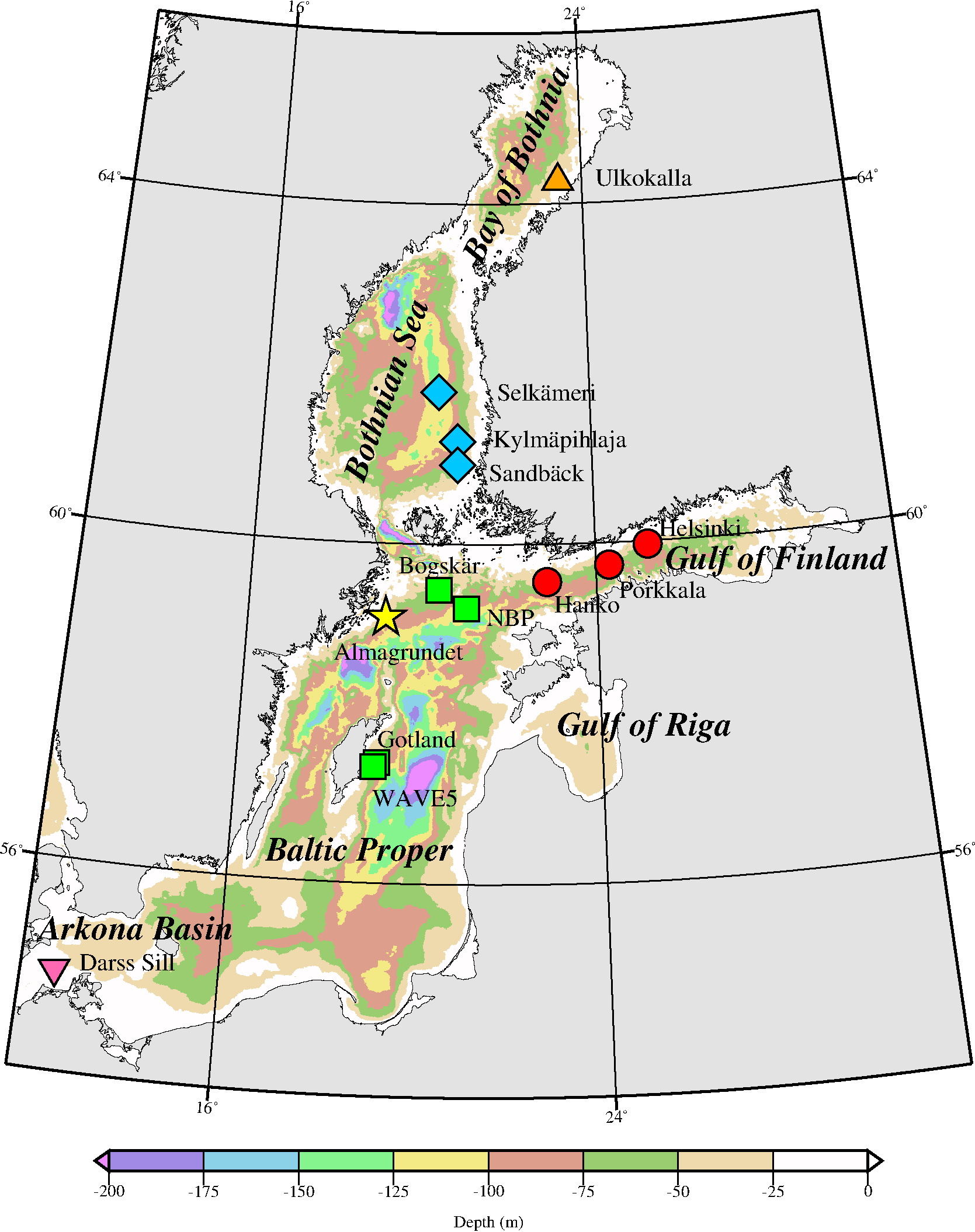}
\caption{The location of the available wave observations. The symbols indicate the grouping used in certain parts of the validation, whereas the colour scale describes the bathymetric data.}
\label{Fig:iow_bal}
\end{center}
\end{figure}

\subsection{Wave buoy data}
We validated the modelled significant wave height ($H_s$) and peak period ($T_p$) against measurements from 13 different locations in the Baltic Sea (Fig. \ref{Fig:iow_bal}). Some of the data originated from short measuring campaigns done at chosen locations ranging from a couple of weeks to several months in time, while some were available from continuous operational measurements conducted at fixed locations year after year. The data were collected every 0.5--3 hours. Because the model parameters were saved once an hour, only coinciding measured values were taken into account. Long time series coincident with the hindcast period are available from Almagrundet (29 years, water depth 30 m), Darss Sill (15 years, water depth 20 m), the Gulf of Finland (14 years, water depth 62 m), Gotland (11 years, water depth 36 m) and the Northern Baltic Proper (10 years, water depth 100 m). The data sets from different institutes will now be briefly introduced.

\subsubsection{FMI}\label{sssec:FMI}
The data originated from wave buoy measurements carried out between 1973 and 2005 by the Finnish Institute of Marine Research (FIMR)\footnote{from 2009 by the Finnish Meteorological Institute (FMI)}. During the 1970s and 1980s, the measurements were made once every three hours with Datawell Waveriders using a 15 minute time series with a sampling time of 0.4 s. The only exception was the wave buoy at Bogsk{\"a}r, which used a ten minute time series once an hour. Starting from the 1990s, the measurements were made with Datawell Directional Waveriders using a 1600 s time series with a sampling time of 0.78 s. One exception was the storm of December 2004, when the data from the operational buoy in the Northern Baltic Proper was retrieved from the on-board data logger. The data logger uses a 1320 s time series to calculate the spectrum. 

During September and October in 2003 an Air-Sea Interaction Spar buoy \citep[ASIS,][]{Graber2000} was moored at the location of the Gotland wave buoy instead of a Directional Waverider. The wave measurements were part of a field study where the University of Miami provided the ASIS buoy \citep{Hogstrom2008}. The difference between the devices should not affect the model validation, since a good agreement between ASIS and Directional Waveriders has been established by \citet{Pettersson2003}.

The time series from Sandb{\"a}ck for the years 1974--75 and from the Gotland wave buoy were transferred 30 minutes forward to obtain a usable number of coinciding points. This did not cause a discrepancy in the Gotland data, since the original time stamp was the starting time for the 30 minute time series. Because of the 15 minute time series used at Sandb{\"a}ck, the change in the time stamps resulted in a discrepancy of 15 minutes. While not ideal, this is an acceptable level of uncertainty in order to be able to also validate the model against these data. 

\subsubsection{SMHI}
The measurement data set near a caisson lighthouse at Almagrundet represents the longest instrumentally measured wave time series in the Baltic Sea. The measurements started in 1978 when SMHI installed an inverted echo-sounder at a depth of roughly 30 m. The devices sampled the water surface position with 10 Hz for about 11 minutes every hour. After some processing involving, e.g. low-pass filtering, the wave spectra was then calculated piece-wise in ten slices. A replacement, but analogous, device was installed in 1992, allowing for more than three years of coinciding measurements with the original device before it was decommissioned in 1995. The wave observations ceased in 2003.

We retrieved the data for this study from the SMHI open data portal. In contrast to the other measurements, the wave heights were not significant wave height $H_s=H_{m_0}$, but $H_{1/3}$, which is defined as the mean value of the highest one-third of the waves. This parameter was estimated from the 10\textsuperscript{th} highest waves, assuming that the waves are Rayleigh distributed \citep[see][]{Broman2006}. The parameters $H_{m_0}$ and $H_{1/3}$ are similar enough to be compared. In deep water for a narrow spectrum, \citet{LH1980} found a best fit of $H_{1/3}=0.925H_{m_0}$. The deep water assumption is valid at Almagrundet for wave periods under roughly 6.5 s. The data did not contain the peak period, only the zero-upcrossing period ($T_z$). as there is no fixed relation between $T_z$ and $T_p$ we did not use period information for this location.

As noted by \citet{Broman2006}, the data set is of varying quality. This is true especially for the data from the new device. We therefore chose not to use the data from the years 1996--97, since they contained many spikes without any apparent physical explanation. In January 1993, the data indicate a constant value for the significant wave height for 104 hours, and these points were removed manually. A more in-depth discussion of these measurements has been provided by \citet{Broman2006}, and \citet{Martensson1987}.

\subsubsection{HZG}
The Darss Sill wave buoy measurements from the Arkona Basin were available for this study. They have been conducted by the Helmholtz Zentrum Geesthacht since 1991 using a Directional Waverider Mk-II that is anchored at a depth of 20 m. The measurement principle of the Mk-II has essentially been described above in Sec. \ref{sssec:FMI}, since it is identical to the Mk-III Directional Waveriders used by FIMR since the 1990s. No peak period ($T_p$) data are currently available from Darss Sill.

\subsection{Altimeter and scatterometer wind data}
We validated the BaltAn65+ surface winds using remotely sensed, quality controlled, satellite radar data, which were available for the years 1992--2005. Quality controls of the satellite wind measurement database include error flags, outlier removal, ice flags and wind sanity checks \citep{Groenewoud2011, Schrama2000, Naeije2008}. In validating the surface wind data of BaltAn65+ we excluded samples closer than roughly 30 km from the coast. Ice and land to sea breezes cause a large scatter and bias in the satellite radar data along the coast, which obfuscates the validation of the model winds on the open sea. Hence, only the Baltic Proper and Bothnian Sea are considered in the comparison with the remotely sensed data; this amounted to some 500,000 altimeter- and 750,000 scatterometer-collocated records. 
\begin{figure}
\begin{center}
\includegraphics[width=9cm]{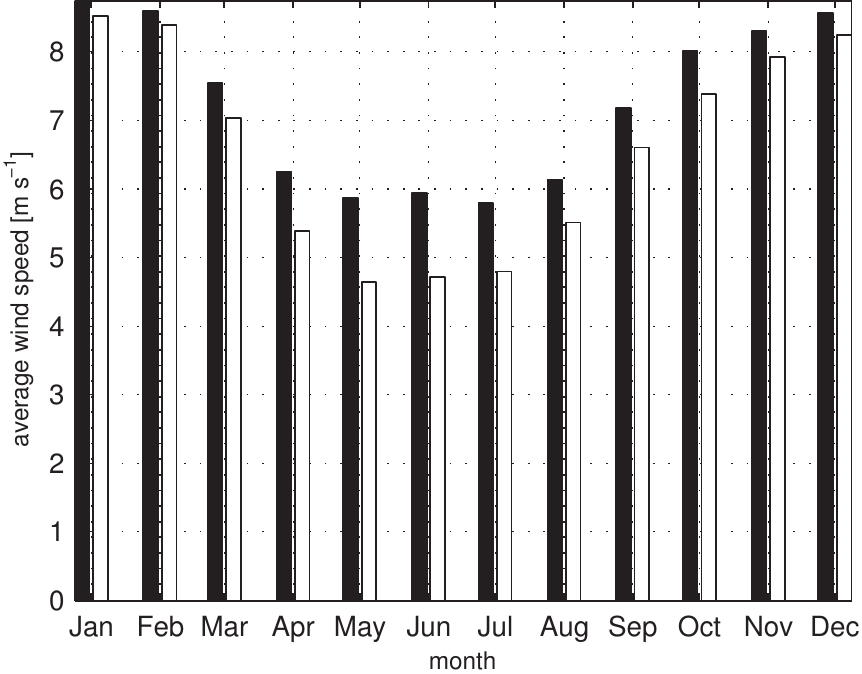}
\caption{The average modelled (black) and observed (white) wind speed per month.}
\label{Fig:wind}
\end{center}
\end{figure}

\section{Model validation}\label{sec:Validation}
 
\subsection{Wind validation}
The correlation between the measured and the modelled wind speed was very good in the Baltic Proper (0.91) and somewhat lower in the Bothnian Sea (0.86). The same finding is also reflected in the RMSE: 0.21 ms$^{-1}$ in the Baltic Proper and 0.26 ms$^{-1}$ in the Bothnian Sea. The seasonal distribution of the average modelled and measured wind speeds (Fig. \ref{Fig:wind}) shows a clear pattern: the modelled wind speed almost matches the measured ones in winter months, but in the summer the hindcast values are higher compared to measurements by 1.1 ms$^{-1}$. This is likely due to the low sea states in the Bothnian Sea when considering that the derivation algorithms for retrieving wind speed from satellites are tuned for the open Ocean \citep{Abdalla2012}. In the Bothnian Sea in summer months, the wind speeds (and sea state) are lower and almost match the backscatter saturation level of the satellite measuring capability. 

Since the data assimilation by \citet{Luhamaa2011} also included coastal observations, it is unlikely that the model really overestimates the wind speed on average. The small negative overall bias in the wave model is also in line with this conclusion (Table \ref{tab:bias_rmse}). While not the sole factor, the overestimation of the forcing wind speed would most likely also be reflected in the wave model.

\begin{table}[ht]
\caption{An overview of the bias and RMSE of the hindcast when compared against all wave measurements. The observations have been grouped according to their respective sub-basin (bolded). The number of coinciding data points and years of operation are also shown for each location.}
\begin{center}
\resizebox{16cm}{!}{
  \begin{tabular}{l c c c c r l}
	Location & \multicolumn{2}{c}{$H_s$ (m)} & \multicolumn{2}{c}{$T_p$ (s)} & Data points & Years\\ \hline
	
	& bias & RMSE & bias & RMSE & & \\   \hline
\multicolumn{1}{l}{\textbf{Baltic Proper}} & \textbf{-0.02} & \textbf{0.25} & \textbf{-0.16} & \textbf{1.05} & \textbf{110220} & \\ \hline
\multicolumn{1}{r}{Northern Baltic Proper} & -0.05 & 0.26 & -0.25 & 1.00 & 44903 & 1996-2005\\ 
\multicolumn{1}{r}{Bogsk{\"a}r} & 0.00 & 0.29 & 0.02 & 1.09 & 14635 & 1982-86\\ 
\multicolumn{1}{r}{Gotland} & 0.01 & 0.22 & -0.13 & 1.08 & 49453 & 1995-2005\\ 
\multicolumn{1}{r}{WAVE5} & 0.00 & 0.21 & -0.21 & 1.20 & 1229 & 2003\\ 
\multicolumn{1}{l}{\textbf{Gulf of Finland}} & \textbf{-0.00} & \textbf{0.20} & \textbf{-0.41} & \textbf{1.31} & \textbf{29130} & \\ \hline
\multicolumn{1}{r}{Helsinki} & 0.00 & 0.20 & -0.44 & 1.36 & 25177 & 1982-85, 90-92, 94 \& 2000-05\\ 
\multicolumn{1}{r}{Porkkala} & -0.02 & 0.19 & -0.30 & 0.99 & 2779 & 1993\\ 
\multicolumn{1}{r}{Hanko} & -0.07 & 0.31 & -0.02 & 0.91 & 1174 & 2001\\ 
\multicolumn{1}{l}{\textbf{Bothnian Sea}} & \textbf{-0.04} & \textbf{0.30} & \textbf{-0.18} & \textbf{0.73} & \textbf{3171} & \\ \hline
\multicolumn{1}{r}{Sandb{\"a}ck} & 0.12 & 0.56 & 0.16 & 0.78 & 176 & 1973-75,1981\\ 
\multicolumn{1}{r}{Kylm{\"a}pihlaja} & 0.01 & 0.25 & -0.10 & 0.72 & 2012 & 1992\\ 
\multicolumn{1}{r}{Selk{\"a}meri} & -0.17 & 0.31 & -0.42 & 0.74 & 983 & 1998-99\\ 
\multicolumn{1}{l}{\textbf{Bay of Bothnia}} & \textbf{-0.08} & \textbf{0.24} & \textbf{-0.16} & \textbf{0.84} & \textbf{1579} & \\ \hline
\multicolumn{1}{r}{Ulkokalla} & -0.08 & 0.24 & -0.16 & 0.84 & 1579 & 1980-1981\\ 
\multicolumn{1}{l}{\textbf{Almagrundet}} & \textbf{0.14} & \textbf{0.31} & \textbf{-} & \textbf{-} & \textbf{115486} & \\ \hline
\multicolumn{1}{r}{Almagrundet} & 0.14 & 0.31 & - & - & 115486 & 1973-1995 \& 1998-2003\\ 
\multicolumn{1}{l}{\textbf{Arkona Basin}} & \textbf{0.09} & \textbf{0.25} & \textbf{-} & \textbf{-} & \textbf{33084} & \\ \hline
\multicolumn{1}{r}{Darss Sill} & 0.09 & 0.25 & - & - & 33084 & 1991-2005\\ 
	
  \end{tabular}}\label{tab:bias_rmse}

\end{center}
\end{table}

\subsection{Wave validation}
General results for an entire basin cannot necessarily be deduced from a few measurement points. Nevertheless, for the purpose of wave validation, the Baltic Sea is divided into five sub-basins: the Bay of Bothnia, the Bothnian Sea, the Baltic Proper, the Gulf of Finland and the Arkona Basin. In addition, Almagrundet is treated separately from the rest of the data from the Baltic Proper because it measured a different parameter (i.e. $H_{1/3}$). The six categories are indicated by the symbols in Fig. \ref{Fig:iow_bal}. The number of data points and years of operation for the different wave measurements are gathered in Table \ref{tab:bias_rmse}, along with the location specific bias and RMSE of the hindcast. The group-specific results will now be discussed.

\subsubsection{The significant wave height}
\begin{figure}
\begin{center}
\includegraphics[width=12cm]{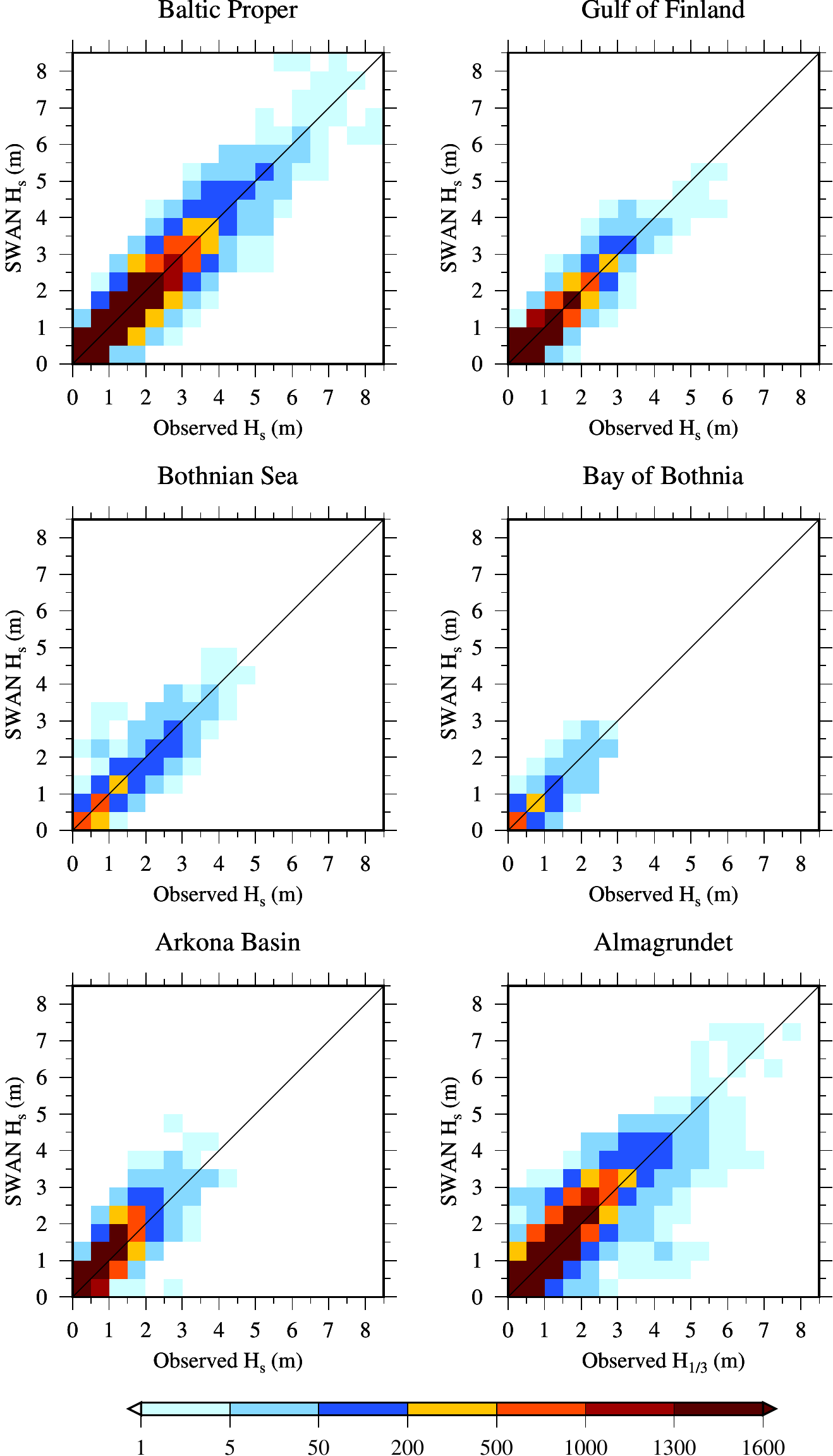}
\caption{Comparison of the significant wave height of the hindcast and the wave measurements from the six regions. The number of coinciding points are shown by the colour scale.}
\label{Fig:Hs_validation}
\end{center}
\end{figure}

The modelled significant wave height compare well to the measurements in all regions, with the exception of Almagrundet (Fig. \ref{Fig:Hs_validation}). The validation was most robust in the Baltic Proper, where 110,220 data points from four locations result in a good match when compared to the hindcast. The model's performance was very good even for the Gotland wave buoy (Table \ref{tab:bias_rmse}), which is located only about $4.5$ km from the small island of {\"O}stergarnsholm near the island of Gotland, in Sweden. As previously noted, we had no instrumental observations with which to validate the hindcast in the eastern or southern Baltic Proper.

The model performed well in the Gulf of Finland despite the challenges imposed by the narrow fetch geometry identified by \citet{Pettersson2010}. The hindcast had a zero bias when compared to the largest data set from the Helsinki wave buoys. A slightly larger bias was found for the most western point (Hanko), although the number of data points was quite low (Table \ref{tab:bias_rmse}).

The hindcast in the Bothnian Sea and the Bay of Bothnia performed somewhat variably (Table \ref{tab:bias_rmse}), but a good overall agreement could still be found (Fig. \ref{Fig:Hs_validation}, middle row). The few severely overestimated points in the Bothnian Sea can be traced to a single event at Sandb{\"a}ck in 1974. However, the good performance of the model at other times and locations suggests that this was an outlier related to a single incident, most likely caused by an overestimation of the wind field. The validation was somewhat limited by the lack of data from the Gulf of Bothnia, but the situation has fortunately been improved in recent years by the addition of FMI's and SMHI's new operational wave buoys in both the Bothnian Sea and the Bay of Bothnia.

In the Arkona Basin, represented by the single point at Darss Sill, the significant wave height had a slight, but systematic overestimation, which was more pronounced with increasing wave heights. The bias was still in the same magnitude as at the other locations, only positive. \citet{Soomere_etal_2012} validated a different wave model (WAM) with a 5.5 km spatial resolution against the Darss Sill measurements using two different wind forcings, finding RMSE values of 0.41 m and 0.72 m and biases of 0.09 m and 0.04 m. Likewise \citet{Siewert2015} compared SWAN to observations from, e.g. Darss Sill, but they did not calculate comparable metrics. In this study, the bias and RMSE at Darss Sill was 0.09 m and 0.25 m respectively (Table \ref{tab:bias_rmse}), suggesting that a higher spatial resolution increases the accuracy of wave model hindcasts even 25 km from the shoreline. However, this finding isn't conclusive because of the different models and wind forcings used in the study. 

The modelled data compared least well to the observations from Almagrundet, which was also the longest data set. We attribute the large scatter to the slightly weaker quality of the observations, since shortcomings in this particular data set were identified both in this study and previously by \citet{Broman2006}. The higher bias at Almagrundet relative to the other locations can to some extent be attributed to the difference in parameters (i.e. $H_{1/3}$ and $H_{m_0}$). However, since a positive bias was also observed at Darss Sill, further research on the reliability of the measured data set is required before any definitive conclusions can be drawn about the performance of SWAN and BaltAn65+ at this location.

\subsubsection{The peak period}

The peak period has been validated for all stations except Almagrundet and Darss Sill, where they were not available.

\begin{figure}
\begin{center}
\includegraphics[width=12cm]{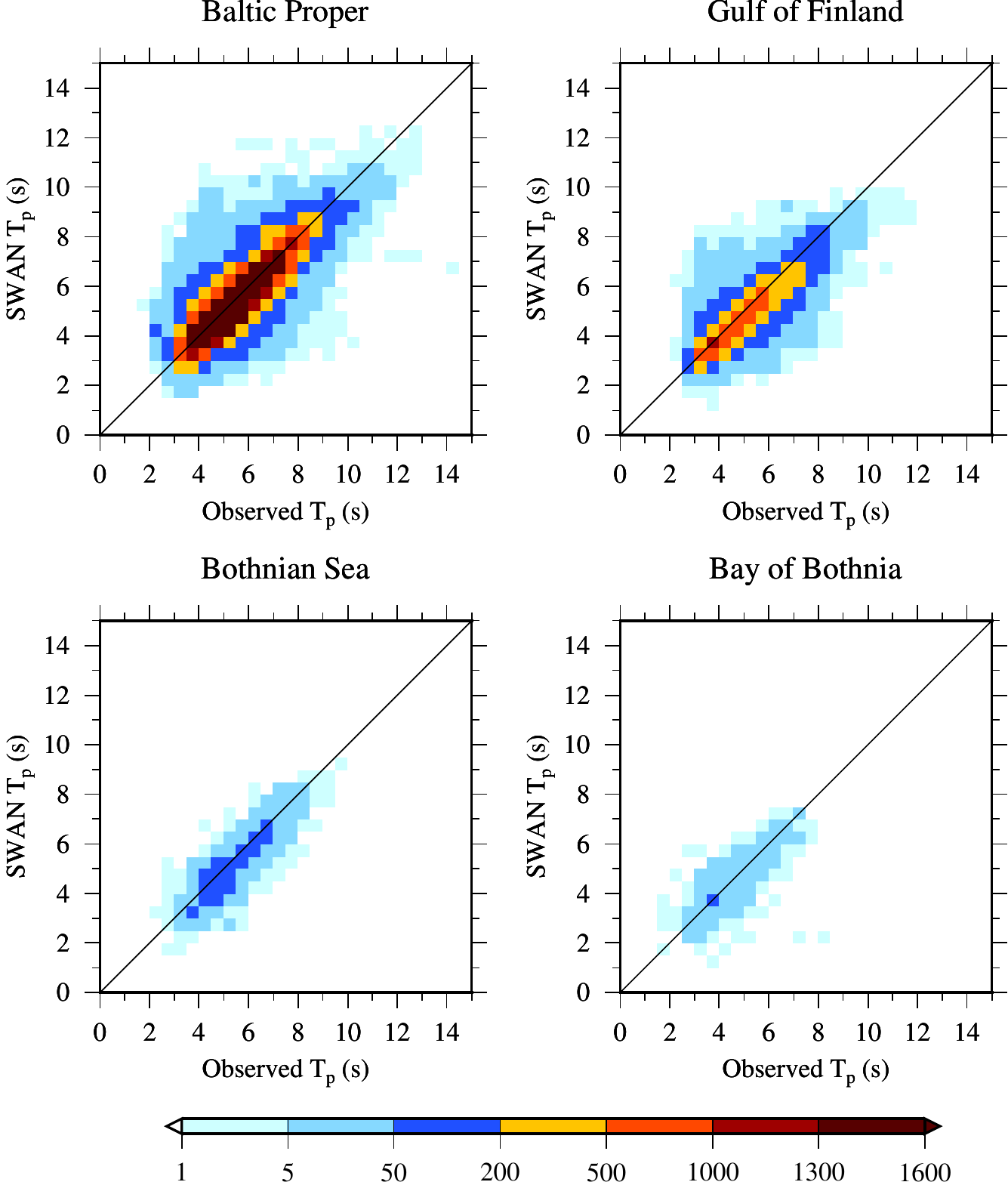}
\caption{Comparison of the peak period of the hindcast and the wave measurements from four regions. Only cases when the observed significant wave height was over 0.5 m are included. The number of coinciding points are shown by the colour scale.}
\label{Fig:Tp_validation}
\end{center}
\end{figure}

Throughout the study, the peak period was modelled slightly less accurately than the significant wave height. Nonetheless, the quality of the hindcast was fair, especially if cases with a significant wave height under 0.5 m were excluded (Fig. \ref{Fig:Tp_validation}). The lack of a well-defined spectral peak for low-energy situations is reflected in a poorly defined peak period in the observations. While the criteria of 0.5 m is admittedly somewhat arbitrary, some threshold should be used in this kind of extensive validation; the peak period cannot be accurately represented in the hindcast in situations when it is not well defined in the first place. For the sake of completeness, the bias and RMSE of the entire data set are also given in Table \ref{tab:bias_rmse}, but the discussion below is limited to cases with a significant wave height over 0.5 m.

The hindcast represents the peak period at the Baltic Proper without any severe bias, but with a clearly larger scatter than in the significant wave height (Fig. \ref{Fig:Tp_validation}, top left). Observed peak periods exceeding 11 s were underestimated by 1.1 s in the hindcast. As illustrated in Fig. \ref{Fig:Tp_validation} (upper right), the highest peak periods of over 10 s at the Helsinki wave buoy were underestimated by roughly 2 s in the hindcast. These cases include the highest sea states, e.g. the measured maximum significant wave height at the Helsinki wave buoy in November 2001 (5.2 m).

In the Bothnian Sea and Bay of Bothnia the scatter was clearly smaller than in the Baltic Proper and the Gulf of Finland (Fig. \ref{Fig:Tp_validation}, bottom). The hindcasts in both areas presented a slight systematic negative bias over the entire range of values.

Overall, while the peak period was fairly represented in a mean sense, its accuracy was still clearly weaker than the significant wave height. This is not surprising as the estimation of the peak frequency is more sensitive to inherent measurement variability than the significant wave height. Large peak periods in particular were underestimated in the hindcast. The accuracy of the peak period will be discussed further in Sect. \ref{sec:comparing_excv}.

\section{Wave field statistics}\label{sec:Wave_stat}

\subsection{Mean statistics}

\begin{figure}
\begin{center}
\includegraphics[width=14cm]{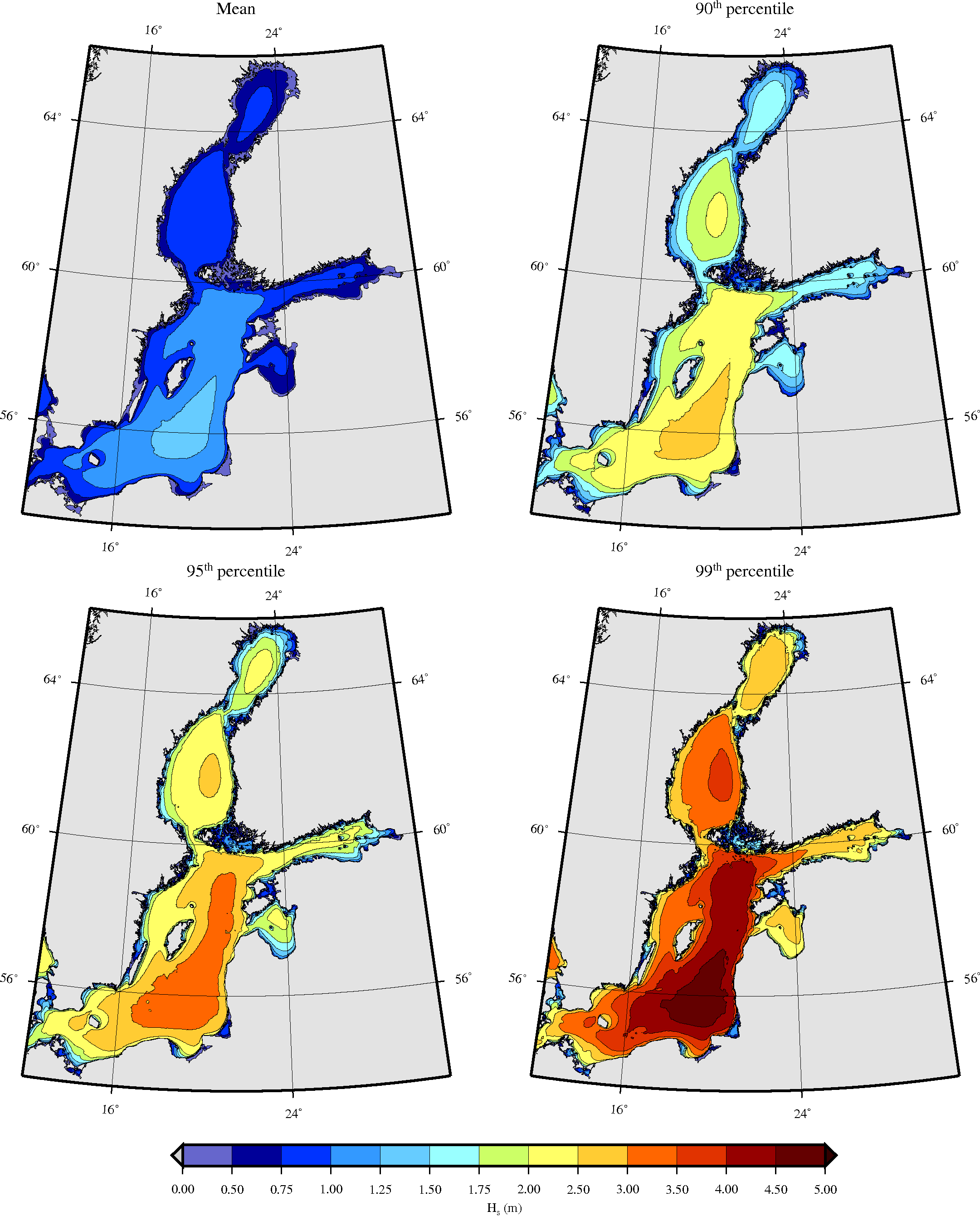}
\caption{Ice-free statistics (Type F in \citet{Tuomi2011}) for the significant wave height. The mean values and the 90\textsuperscript{th}, 95\textsuperscript{th} and 99\textsuperscript{th} percentiles are given.}
\label{Fig:Hs_noice}
\end{center}
\end{figure}

We calculated wave field statistics for the entire Baltic Sea for the ice-free periods \citep[Type F in][]{Tuomi2011} from the entire 41-year model hindcast. The ice-free period is defined separately for each grid point and is simply all the times when the grid point is not covered in ice. As expected, the highest average significant wave height occurred in the Baltic Proper, while the lowest average values could be found in the small sub-basins of the Bay of Bothnia and Gulf of Riga (Fig. \ref{Fig:Hs_noice}). The hindcast placed the harshest wave climate in the southern part of the Baltic Proper; no wave measurements are available from this region. The wave climate in the Arkona Basin, where Darss Sill is located (see Fig. \ref{Fig:iow_bal}), varied significantly over relatively short distances, which is related to strong variations in upwind fetch restrictions. 

The hindcast is able to resolve some near shore dynamics in more detail than the previous 6 nmi hindcast of \citet{Tuomi2011}. The south-eastern part of the GoF has lower significant wave heights than the northern part of the gulf (Fig. \ref{Fig:Hs_noice}). However, the results for the Archipelago Sea cannot be considered reliable, even though its wave field is technically resolved by the model. Special modelling techniques are required to simulate the islands effect on the wave field \citep{Tuomi2014}.

\begin{figure}
\begin{center}
\includegraphics[width=14cm]{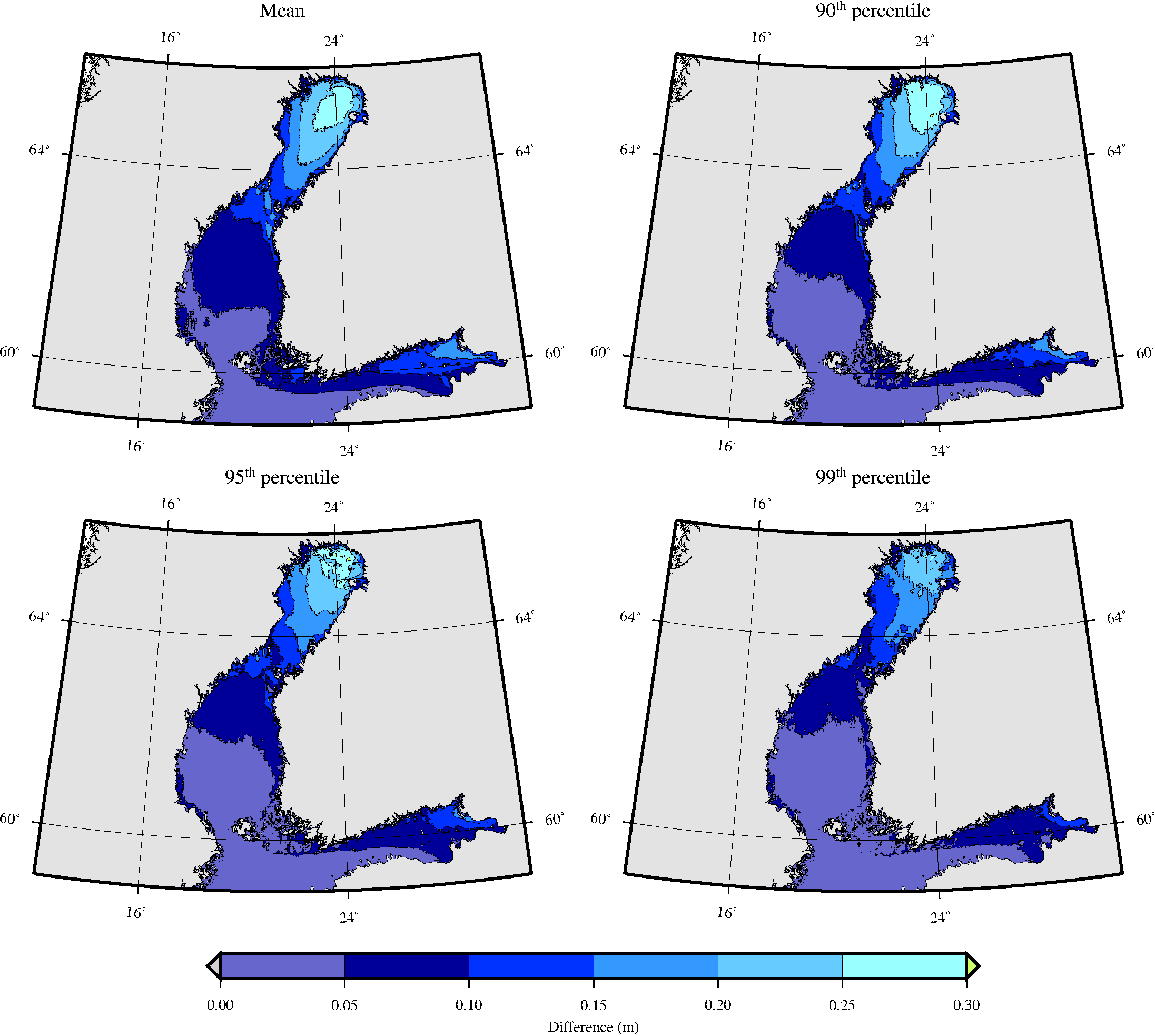}
\caption{Difference between ice-free and ice-included statistics (Type F and I in \citet{Tuomi2011}) for the significant wave height. The mean values and the 90\textsuperscript{th}, 95\textsuperscript{th} and 99\textsuperscript{th} percentiles are given.}
\label{Fig:Hs_diff}
\end{center}
\end{figure}

The highest wave events (90\textsuperscript{th}, 95\textsuperscript{th} and 99\textsuperscript{th} percentiles) particularly occurred in the southern part of the Baltic Proper. This is somewhat in contrast to the findings presented by \citet{Tuomi2011}, who found the highest wave events in the Northern Baltic Proper (NBP). These differences cannot be explained by the use of ice-included statistics in the study by \citet{Tuomi2011}, since the difference is small in the Baltic Proper (Fig. \ref{Fig:Hs_diff}). This geographic difference at higher percentiles can partially be explained by the shorter time period used by the study of \citet{Tuomi2011}, which was from 2001--2007.

The difference between the ice-free and ice-included statistics for the mean significant wave height were large in the Bay of Bothnia, where they differed by 30\% (Fig. \ref{Fig:Hs_diff}). The differences between these two types of statistics were quite large also in the eastern parts of the Gulf of Finland. Below a latitude of 59.5\textdegree\ the difference was smaller than 0.05 m almost everywhere (not shown).

The results for the mean values of the significant wave heights agreed with those provided by \citet{Tuomi2011} after accounting for the difference between the ice-free and ice-included statistics. In a study by \citet{Raamet2010}, the Baltic Sea wave field was simulated for the years 1970--2007 without accounting for the ice. These so-called hypothetical no-ice statistics \citep[Type N in][]{Tuomi2011} produces larger values than the ice-free statistics, since the fetch during any winter storm is never limited by the ice cover. The significant wave height presented here is still up to 0.40 m higher in the Baltic Proper compared to the findings provided by \citet{Raamet2010}. This can mainly be attributed to a difference in the wind forcing. \citet{Raamet2010} used geostrophic winds with a resolution of nearly 111 km, which the authors found to result in a systematic overall underestimation of waves by 0.20 m in the Baltic Proper.

\begin{figure}
\begin{center}
\includegraphics[width=14cm]{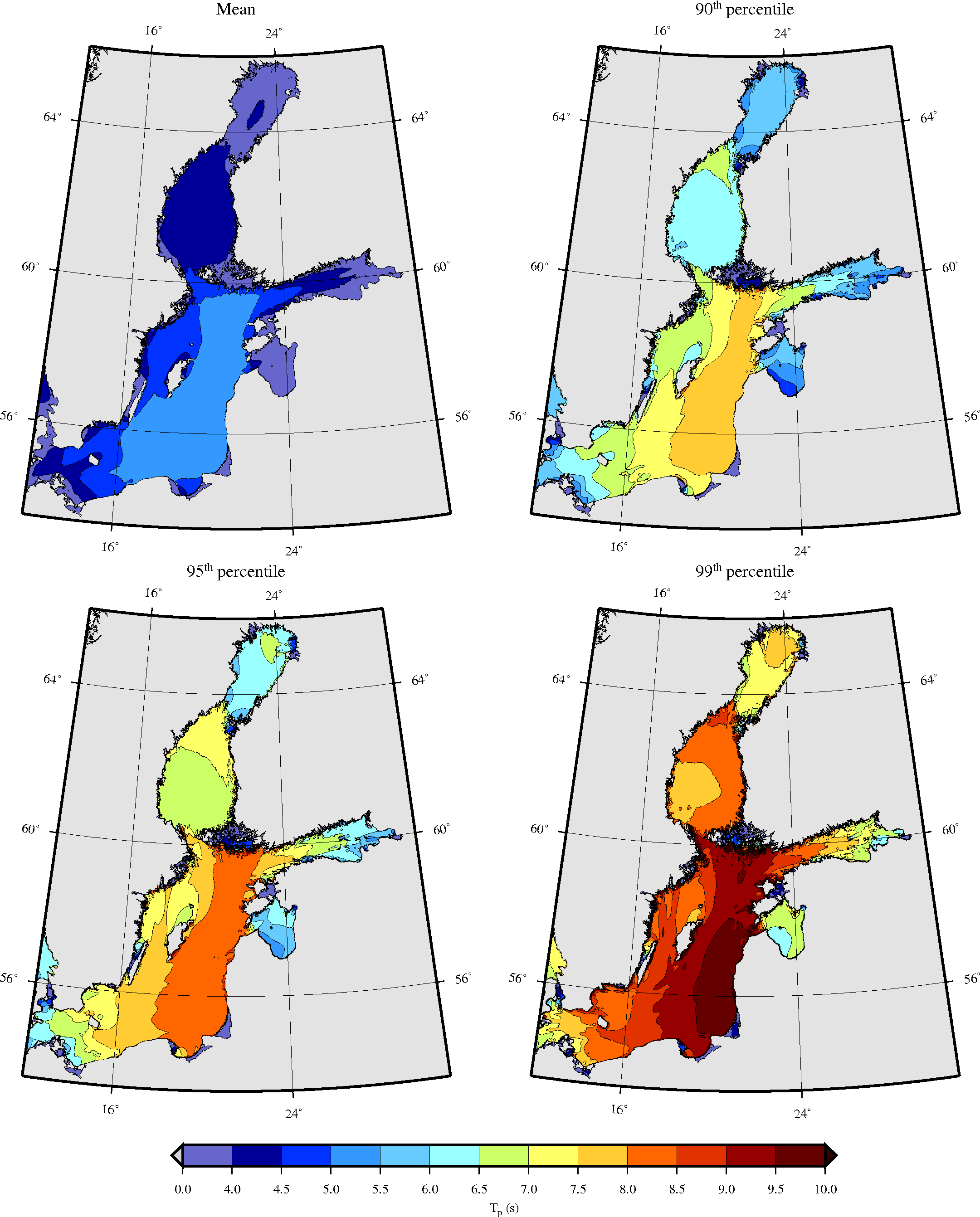}
\caption{Ice-free statistics (Type F in \citet{Tuomi2011}) for the peak period. The mean values and the 90\textsuperscript{th}, 95\textsuperscript{th} and 99\textsuperscript{th} percentiles are given.}
\label{Fig:Tp_noice}
\end{center}
\end{figure}

Similarly to the significant wave height, the highest mean and percentile values for the peak period were found in the Baltic Proper, where the mean values are 5--6 s and the 99\textsuperscript{th} percentile reaches 10 s (Fig. \ref{Fig:Tp_noice}). In the narrow Gulf of Finland the peak period increased up to 9 s during storms, but remained under 5 s on average. The mean peak periods in the Gulf of Riga were between 3 and 4 s, since the propagation of longer waves from the Baltic Proper to the Gulf of Riga is restricted as they are refracted to the sides of the entrance channel connecting the both sea areas. The peak period statistics for the Bothnian Sea resembled those for the Gulf of Finland, while the results show slightly lower values for the Bay of Bothnia (Fig. \ref{Fig:Tp_noice}).

\begin{figure}
\begin{center}
\includegraphics[width=12cm]{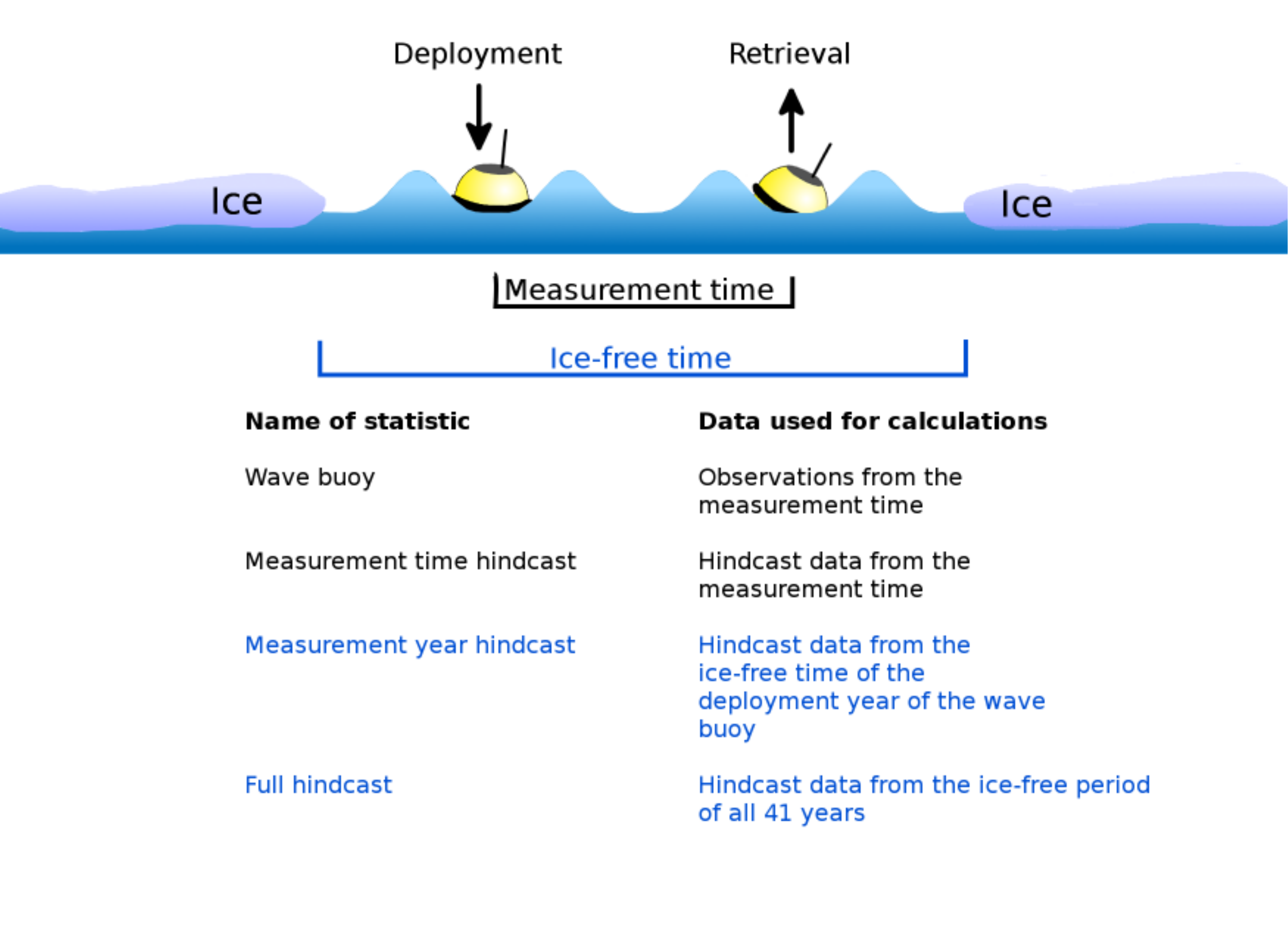}
\caption{A schematic picture illustrating the different time periods and types of statistics. The black colour indicates the restriction to the actual time the wave buoy was deployed ("measurement time"). The blue colour indicates the use of all available modelled wave data ("ice-free time") either for some specific years or the full 41-year hindcast.}
\label{Fig:Schematic}
\end{center}
\end{figure}

\subsection{Extreme events}\label{sec:extreme}
For the purpose of this study, the significant wave height was considered extreme if it exceeded the 99.9\textsuperscript{th} percentile of the significant wave height occurring anywhere in the Baltic Sea between 1965 and 2005. We identified these events simply by registering the maximum significant wave height in the Baltic Sea for each model output time step and calculating the 99.9\textsuperscript{th} percentile, which was 6.9 m. For the sake of convenience, we analysed wave events with a significant wave height of at least 7 m. Using this definition, we identified 45 unique extreme events occurring at different locations during the 41-year hindcast period. Twelve of these events had a maximum significant wave height of over 8 m, with half exceeding 9 m. A significant wave height of over 10 m occurred only once.

All 45 events occurred between September and April, while November--January contained 84 \% of all extreme wave events. Storms in November and December are slightly more extreme in terms of maximum significant wave height when compared to storms in January. The average exceedance times for a significant wave height of over 7 m for December and January were comparable (9 h and 10 h respectively). The severest storm events follow the 99\textsuperscript{th} percentile and were, therefore, found in the southern Baltic Proper (Fig. \ref{Fig:Hs_noice}). In our hindcast, all wave events over 9 m were generated by cyclones passing north from the island of Gotland. The mean wave direction during these storms was not aligned with the strongest winds, but was instead aligned with the long westerly fetch in the southern Baltic Proper due to slanting fetch effects \citep[e.g.][]{Donelan1985, Pettersson2010}. The maximum significant wave height for, e.g. the storm Gudrun in January 2005 was 9.2 m, which is in line with earlier findings \citep{Soomere2008, Tuomi2011}. Nevertheless, in contrast to our hindcast, previous studies have found the maximum significant wave height during Gudrun to be in the Northern Baltic Proper.

The overall maximum significant wave height in the modelled data set reached 10.1 m in the southern Baltic Proper (4 December, 1999). In the Bothnian Sea, the maximum modelled wave height was 7.3 m (19 December 2001), while it was 6.8 m at the entrance to the Gulf of Finland (31 December 1983) and 6.0 m in the Bay of Bothnia (19 December 1992). In the Gulf of Riga, a maximum significant wave height of 6.2 m was modelled as having occurred on 2 November 1969, when wind gusts were close to 50 ms$^{-1}$ \citep{Tarand_etal2013}.

An significant wave height of 8.2 m was measured by the NBP wave buoy in 2004, and this value has not been exceeded since \citep{Bjorkqvist2017b}. The maximum modelled value at the wave buoy in our data set was 8.5 m (in 2002). At the GoF wave buoy a maximum value of 5.2 m has been reached in 2001 \citep[and again in 2012,][]{Pettersson_HELCOM2013}, while the modelled maximum is 4.9 m (in 1986). Also the modelled maximum at the Gotland wave buoy (6.2 m in 1984) is comparable with the measured maximum of 5.6 m, which was recorded in 2017 \citep{Bjorkqvist2017b}. These modelled and measured maximum values form a consistent description of the highest Baltic Sea wave heights.

Overall, the peak periods at instances when the significant wave height exceeded 7 m remained between 9.5 s and 13.6 s. The peak period at the moment of the maximum significant wave height (10.1 m) was 13.4 s. In the Gulf of Finland, the maximum peak periods for wave events over 5 m were up to 10.8 s. This was higher than for the Gulf of Riga and the Bay of Bothnia, where the maximum peak periods did not exceed 9 s. In the Bothnian Sea, the highest peak periods reached 11.2 s in the case of the highest waves.

\section{Estimating exceedance values}\label{sec:Comp}

Determining exceedance values is an integral part of different risk assessment scenarios. The most reliable way to determine return values involves using decades-long continuous measurement time series. These are, however, sparse. Instead, we will determine exceedance values using the long measurement time series from the NBP and the Helsinki wave buoy in the Gulf of Finland, as well as data from the 41-year hindcast. We will also study the impact of using only a shorter subset of the hindcast. Limiting the calculation to the actual time the wave buoy was deployed will quantify how the time that is left unmeasured affects the statistics.

\subsection{Calculation of exceedance values}
Calculating the exceedance values from the entire 41-year hindcast was a straightforward process. We formed a cumulative probability distribution, which gave the exceedance frequencies after multiplying the probabilities by the number of events in one year. The seasonal ice cover did not affect the normalisation, since the maximum value of the time series corresponded with approximately one event in 41 years, regardless of the extent of the ice cover. One wave event was defined as the significant wave height measured for a period of 30 minutes. To be comparable with the wave buoy measurements, the model data were therefore interpolated to 30 minute values.

If wave buoy measurements were not available for the entire open water period, an alternative normalisation had to be performed. The main assumption was that the wave climate during the entire open water period is similar to the time of the measurements. If observations were available for, e.g. only half of the open water period, the exceedance frequency of the measured maximum was two events per year. The normalisation could not include the ice-covered period, which we therefore estimated based on the ice model. A schematic figure of the different relevant time periods can be found in Fig. \ref{Fig:Schematic}. 

\subsection{Comparing measured and modelled exceedance values}\label{sec:comparing_excv}
We calculated exceedance values from the model hindcast using $i)$ the entire 41-year data set and $ii)$ only the calendar years when a wave buoy was deployed. Henceforth, these types of data will be referred to as the full hindcast and deployment year hindcast. Statistics using only observational data will be called wave buoy statistics (Fig. \ref{Fig:Schematic}). 

As the measurements made during one year should represent the wave conditions for that year in some meaningful way, we limited the comparison to years with more than 40 days of observations. This threshold was implemented by excluding years with less than 1920 observations. At the NBP, the observations ranged from 50 to 260 days per year, and we could therefore use the entire data set (years 1996--2005). For Helsinki, we limited ourselves to the years 2001--2005 when observations were available for anywhere between 87 and 200 days per year. 

For Helsinki, the full 41-year hindcast resulted in slightly larger exceedance values of significant wave height compared to the five year wave buoy statistics (Fig. \ref{Fig:RV_cum_Hs}, right). The only exception was the measured maximum (5.2 m), which was underestimated by SWAN. This underestimation is probably explained by the coarse temporal resolution of the wind forcing (6 h), since there was a rapid wave growth and decay of over 1 m in just three hours (see Fig. \ref{Fig:app_timeseries} in the Appendix). However, intense storms can be underestimated by wave models even though the temporal resolution of the wind forcing is as high as 1 h \citep{vanVledderAkpinar2015, Bjorkqvist2017b}.

At the NBP the full hindcast predicted lower exceedance values compared to the wave buoy statistics for exceedance frequencies greater than roughly once per year (Fig. \ref{Fig:RV_cum_Hs}, left).

Figure \ref{Fig:RV_cum_Hs} (left) illustrates the fact that the 1996--2005 measurement year hindcast (red) and the wave buoy statistics (black dots) yielded quite similar results at the NBP. A similar situation can also be observed for the 2001--2005 measurement year hindcast at Helsinki (Fig. \ref{Fig:RV_cum_Hs}, right). By comparing the measurement year hindcasts with the full hindcast it is evident that a 40-year event was captured by the 10-year hindcast at the NBP. In contrast, such a rare event was not included in the five-year hindcast at Helsinki. The different results when using only a subset of the hindcast can mainly be caused by two things: 1) the wave climate during the last ten (or five) years was significantly different than during the preceding 30 years, or 2) even a ten year measurement data set is inadequate for accurately representing the wave climate. Both can, of course, be true simultaneously. 

The exceedance values for the peak period at the NBP calculated from the hindcast were systematically smaller compared to the observations for exceedance frequencies of less than about 300 times per year (Fig. \ref{Fig:RV_cum_Tp}, left). Nevertheless, the difference was under 1 s for both the full hindcast (blue) and the 1996--2005 measurement year hindcast (red). The results from the 41- and ten-year hindcasts were in good accord when accounting for the effect the length of the time series had on the rare exceedance frequencies.

The hindcast at Helsinki produced exceedance values roughly 2 s lower than those determined from the measurements, even for exceedance frequencies of roughly 100 times per year (Fig. \ref{Fig:RV_cum_Tp}, right). The full hindcast (blue) and the 2001--2005 measurement year hindcast (red) yielded practically identical results for exceedance frequencies of more than once per year, but differed for rarer cases.

The low sea states ($H_s<0.5$ m) accounted for 80\% of the cases when a peak period of more than 10 s was observed at the Helsinki wave buoy. For these time, the discrepancy between the measured and modelled peak period at the Helsinki wave buoy can largely be attributed to ship wakes. The period of the ship wakes can exceed 10 s \citep[e.g.][]{Erm_etal2009} and they showed up as visible narrow-banded peaks below 0.2 Hz in the wave buoy spectra, while they were completely missing from the spectra modelled by SWAN. 

Figure \ref{Fig:Tp_validation} (top right) shows that the highest peak periods were underestimated even when the significant wave height exceeded 0.5 m. For these higher sea states, e.g. during the measured $H_s$ maximum at the Helsinki wave buoy in November 2001, the growth of the wave spectra in the hindcast was restricted in the wave model. It is unlikely that the reason is solely an inaccurate forcing wind field, since the variance density at the higher frequencies above 0.15 Hz were modelled quite well. The main discrepancy had to do with the fact that the modelled spectra were clearly less peaked, having roughly four times less energy at the lowest frequencies (Fig. \ref{Fig:app_4spec}). 

The longest fetch in the direction of the Gulf of Finland continues across the Baltic Proper and is of the magnitude of 450 km (Fig. \ref{Fig:iow_bal}). The wind speed during the storm in November 2001 was roughly 20 ms$^{-1}$ according to measurements from the centre of the Gulf of Finland. Thus, this combination would result in a peak period approximately 11.5 s when using the wave growth relations proposed by \citet{KahmaCalkoen1992}, which matches the maximum peak period measured by the wave buoy (11.6 s). This indicates that the growth of the longest fetch components were underestimated, or that they were misdirected.

A comparison to a wave model run without depth-induced wave refraction (see the Appendix) indicated that the discrepancy was, at least partially, related to a too strong refraction of the longer wave periods towards the shores of this narrow gulf. The estimate for the shift in the phase speed ($c_{\theta}$) have been updated already in SWAN 41.01AB and now uses the phase velocity gradients instead of the previous depth gradients. This should mitigate the excessive refraction that has previously been identified in SWAN when using coarse grids in areas with steep depth gradients. A higher resolution bathymetry alleviates the problem slightly (see the Appendix). Exploring a refraction limiter could artificially solve the problem of the underestimation of the peak period, but in narrow fetch situations also the source terms like whitecapping dissipation or non-linear four-wave interactions affect the amount of low-frequency energy \citep[e.g.][]{Pettersson2010}.

Although the modelled peak period can be used to calculate mean statistics, it is not suitable for extreme value analysis because of the negative bias evident in the highest values. One possibility could be to use a more stable characterisation for the wave period, such as the mean wave period $T_{m}$, but the underestimation that occurred during the high sea states would surely have affected also this parameter.

\subsection{The effect of measurement gaps}
Wave buoys seldom collect data during the entire ice-free time, since they often have to be removed for maintenance in time to avoid any damage by freezing, or whenever operationally possible. To quantify the effect of these gaps on the representability of wave buoy observations, we calculated the measurement time hindcasts in addition to the measurement year hindcasts (see Fig. \ref{Fig:Schematic} for definitions). If the wave buoy is removed before an exceptional winter storm, that event will be completely missing from the measurement time statistics. In addition to smaller gaps caused by disruptions in the data transmission, this open-water time before the deployment (or after the recovery) of the wave buoy is the only difference between the measurement time and measurement year statistics calculated from the hindcast. The gaps were typically from December or January until May at the Helsinki wave buoy. These gaps were somewhat shorter at the NBP wave buoy, with an average length of slightly under three months, also typically ending in May. The NBP wave buoy was deployed year round in 2000, 2001 and 2004.

\begin{table}[ht]
\caption{Ice-free time statistics of the wave field parameters at the NBP wave buoy and the Helsinki wave buoy. The definitions for the different time periods used to calculate the statistics can be found in Fig. \ref{Fig:Schematic}.}\label{tab:stat}
\begin{center}
\resizebox{12cm}{!}{
  \begin{tabular}{l c c c c r l}
	 & \multicolumn{2}{c}{Mean} & \multicolumn{2}{c}{95\textsuperscript{th} percentile} & \multicolumn{2}{c}{99\textsuperscript{th} percentile}\\ \hline
	
	 & $H_s (m)$ & $T_p (s)$ & $H_s (m)$ & $T_p (s)$ & $H_s (m)$ & $T_p (s)$ \\   \hline
\multicolumn{1}{l}{\textbf{Northern Baltic Proper}} &  &  &  & &  & \\ \hline
\multicolumn{1}{l}{Hindcast (full)} & 1.20 & 5.2 & 3.00 & 8.2 & 4.22 & 9.3\\ 
\multicolumn{1}{l}{Hindcast (meas. years)} & 1.16 & 5.1 & 2.89 & 8.1 & 4.25 & 9.3\\ 
\multicolumn{1}{l}{Hindcast (meas. time)} & 1.18 & 5.2 & 2.91 & 8.1 & 4.22 & 9.3\\ 
\multicolumn{1}{l}{Wave buoy} & 1.24 & 5.4 & 2.93 & 8.2 & 4.24 & 9.7\\ \hline 
\multicolumn{1}{l}{\textbf{Helsinki}} &  &  &  & &  & \\ \hline
\multicolumn{1}{l}{Hindcast (full)} & 0.84 & 4.4 & 2.07 & 7.0 & 2.90 & 8.1\\ 
\multicolumn{1}{l}{Hindcast (meas. years)} & 0.78 & 4.3 & 1.95 & 6.9 & 2.70 & 8.1\\ 
\multicolumn{1}{l}{Hindcast (meas. time)} & 0.81 & 4.4 & 1.96 & 6.8 & 2.70 & 8.0\\ 
\multicolumn{1}{l}{Wave buoy} & 0.84 & 4.9 & 2.01 & 7.5 & 2.75 & 8.9\\ 
\hline 
  \end{tabular}}

\end{center}
\end{table}

\begin{figure}
\begin{center}
\includegraphics[width=14cm]{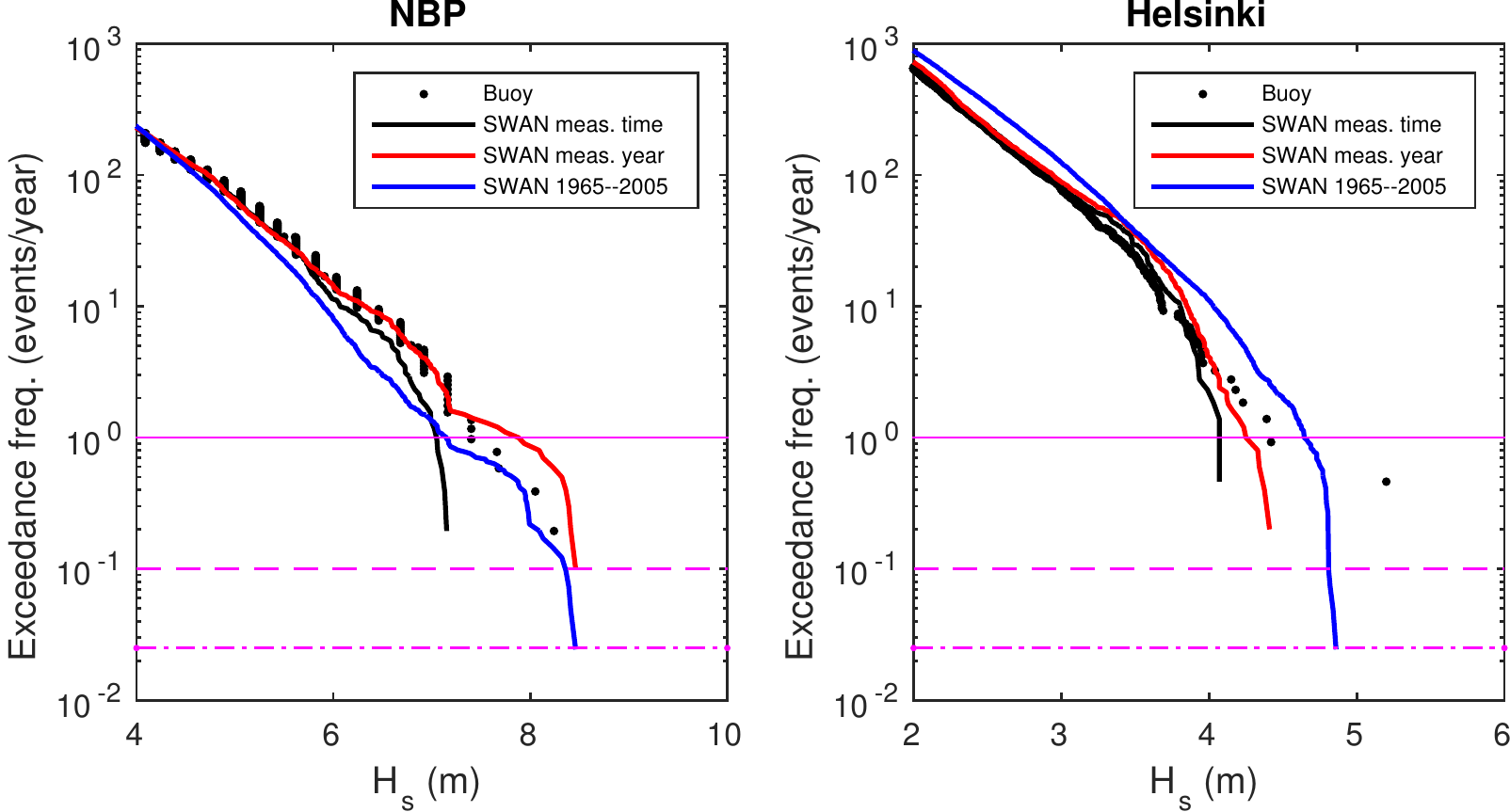}
\caption{The significant wave height at NBP (left) and Helsinki (right). The measurement time hindcast (black) is restricted to values coinciding with the wave buoy observations (black dots). The measurement year hindcast (red) covers the full deployment years of the wave buoy. The exceedance frequencies of once per 1, 10 and 40 years are given by the solid, dashed, and dashed-dotted horisontal lines. One event in this figure is that of the significant wave height sustained for 30 minutes.}\label{Fig:RV_cum_Hs}

\end{center}
\end{figure}

The Helsinki data set showed only a slight difference in exceedance frequencies lower than once per year when using the measurement time hindcast (black) as opposed to using the 2001--2005 measurement year hindcast (red) (Fig. \ref{Fig:RV_cum_Hs}, right). This indicates that the information lost by the measurement gaps was quite small.

In the NBP data set there was a remarkable difference when we restricted ourselves to the measurement time hindcast (black) instead of using the 1996--2005 measurement year hindcast (red). The full ten-year hindcast data match the observations better in comparison to only the hindcast values coinciding with the measurements (Fig. \ref{Fig:RV_cum_Hs}, left). While somewhat surprising, this finding is well explained by two storm events. The storm Rafael in 2004 occurred during the time when the wave buoy was deployed, but was underestimated by up to 1.6 m by the hindcast as a result of a too weak forcing wind speed. This explains the difference between the wave buoy statistics (black dots) and the statistics from the measurement time hindcast (black). The ten-year hindcast again captured a storm in February 2002 after the wave buoy had been retrieved, which explains the difference between the measurement time hindcast (black) and the 1996--2005 measurement year hindcast (red).

\begin{figure}
\begin{center}
\includegraphics[width=14cm]{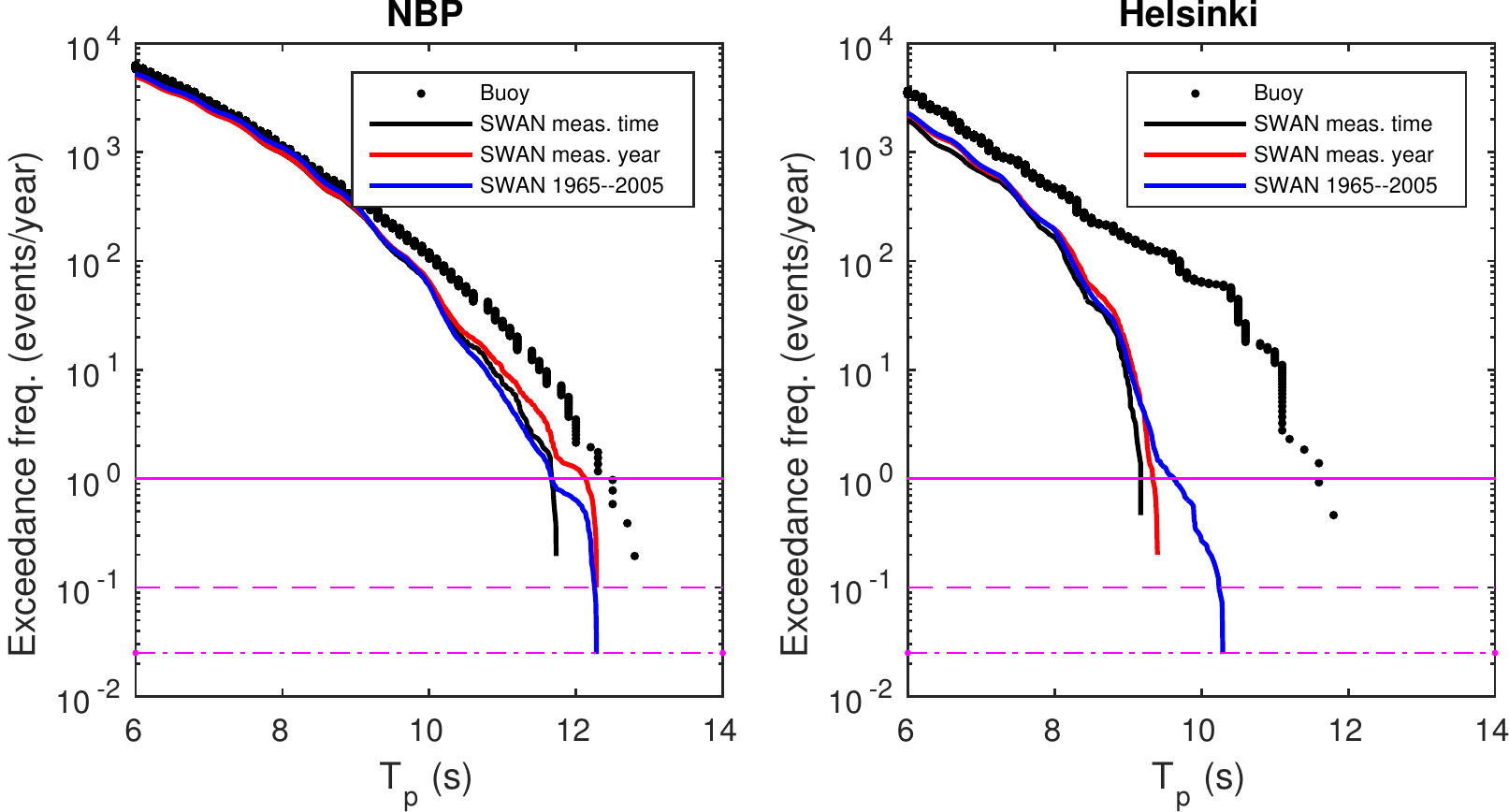}
\caption{The peak period at NBP (left) and Helsinki (right). The measurement time hindcast (black) is restricted to values coinciding with the wave buoy observations (black dots). The measurement year hindcast (red) covers the full deployment years of the wave buoy. The exceedance frequencies of once per 1, 10 and 40 years are given by the solid, dashed, and dashed-dotted horisontal lines. One event in this figure is that of the peak period sustained for 30 minutes.}\label{Fig:RV_cum_Tp}
\end{center}
\end{figure}

The same storm event from 2002 was also reflected in the peak period at the NBP (Fig. \ref{Fig:RV_cum_Tp}, left), resulting in a difference of 0.5 s for rare exceedance frequencies. The measurement gaps had practically no influence on the exceedance values of the peak period at Helsinki (Fig. \ref{Fig:RV_cum_Tp}, right).

The impact of the measurement gaps was not determined by their mean length (175 days at NBP and 187 days at Helsinki). However, it was crucial to capture all storms, which are most prevalent between September and April. The mean values and the 95\textsuperscript{th} and 99\textsuperscript{th} percentiles were affected by less than 5\% by the measurement gaps (Table \ref{tab:stat}).

\subsection{Exceedance values for the Baltic Sea}

\begin{figure}
\begin{center}
\includegraphics[width=14cm]{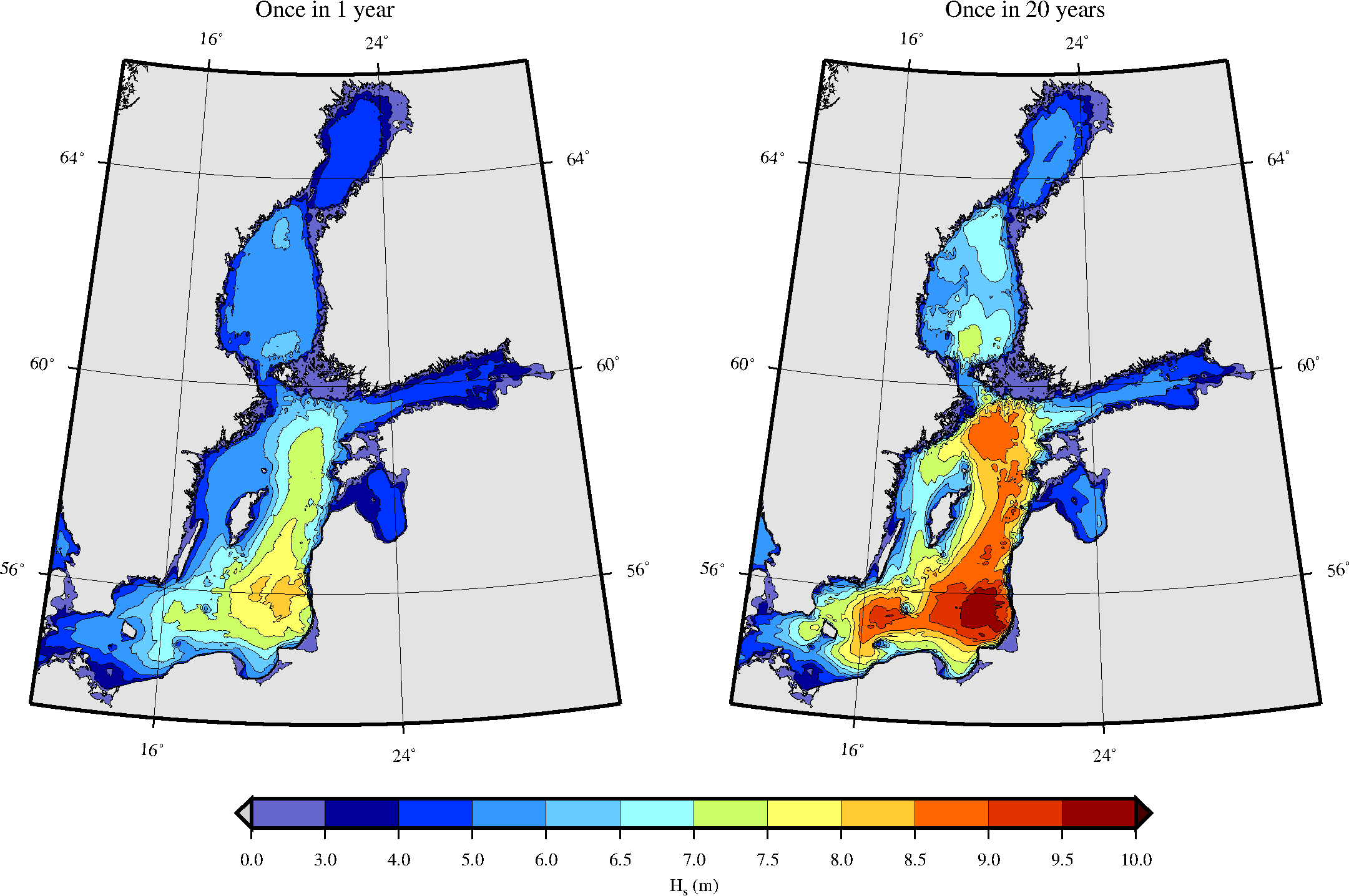}
\caption{The significant wave height for exceedance frequencies of once per year (a) and once per 20 years (b). One event in this figure is that of the significant wave height sustained for one hour. }
\label{Fig:Hs_exceedande}
\end{center}
\end{figure}

Since the hindcast was found to be sufficiently accurate, we calculated the wave heights corresponding to exceedance frequencies of once in 1 year and once in 20 years for the entire Baltic Sea (Fig. \ref{Fig:Hs_exceedande}). For computational reasons, the gridded hindcast results were not interpolated to a 30 minute time step. The values in Fig. \ref{Fig:Hs_exceedande} can therefore directly be read as hours in the given time period. This interpretation coincides with the exceedance time (ET) statistic in \citet{Tuomi2011}. 

In the southern Baltic Proper, the significant wave height exceeded 8 m on average once per year (Fig. \ref{Fig:Hs_exceedande}, left), while the corresponding value in the northern part of the Baltic Proper was roughly 7 m. The Bothnian Sea had an exceedance value of over 6 m only in small areas in the very southern and northern parts of the basin. This was in contrast to the mean and 90\textsuperscript{th} -- 99\textsuperscript{th} percentiles, which reached their highest values in the middle of the Bothnian Sea. The smaller basins (Gulf of Riga, Gulf of Finland and Bay of Bothnia) had similar one year exceedance values of less than 5 m.

The exceedance frequency of once in 20 years (Fig. \ref{Fig:Hs_exceedande}, right) is equivalent to two hours in 40 years. In practice, this statistic therefore captured the harshest conditions in the 41-year hindcast. The exceedance value was 8.5 m in most parts of the Baltic Proper, nearing 10 m in the southern part of the basin. In the Bothnian Sea, the largest values of over 7 m were found in the south. The three smaller basins again exhibited similar properties, with exceedance values of at most 6 m.

\section{Summary and conclusions}\label{sec:Conclusions}
We used a 41-year wave hindcast by SWAN forced with reanalysed BaltAn65+ winds for the time period 1965--2005 to study the wave climate of the Baltic Sea. The hindcast was validated against decades of in-situ wave measurements from 13 locations. We compared ice-included and ice-free statistics, and calculated exceedance values obtained from the wave buoy measurements as well as from the full 41-year hindcast. In addition, we studied the hindcast restricted to only the calendar years in which a wave buoy was deployed and quantified the effect of measurement gaps by just using hindcast values that coincided with the measurement points. See Fig. \ref{Fig:Schematic} for an overview of the different types of statistics.

We found that the 41-year hindcast gave an accurate description of the significant wave height in the Baltic Sea. Including the ice time data when calculating the statistics reduced the mean significant wave height by 30\% in the Bay of Bothnia, where the difference between the ice-free and ice-included statistics was up to 0.3 m. The difference remained under 0.05 m below a latitude of 59.5\textdegree. All of the 45 modelled extreme events ($H_s \geq 7$ m) occurred between September and April, while 84\% of the events occurred between November and January. For the highest observed peak periods, the hindcast showed a systematic negative bias. This underestimation was especially prevalent in the Gulf of Finland, which has a narrow fetch geometry, but the results are still fairly accurate in a mean sense across the board (Fig. \ref{Fig:Tp_validation}, Tables \ref{tab:bias_rmse}, \ref{tab:stat}).

Data lost by measurement gaps before and after the ice time can impact the exceedance values of the significant wave height by up to 20\% (Fig. \ref{Fig:RV_cum_Hs}). However, the impact below the 99\textsuperscript{th} percentiles was under 5\% in the NBP and the Gulf of Finland (Table \ref{tab:stat}). In conclusion:

\begin{itemize}

\item The hindcast compared well to wave observations as close as 8 km from the shore and resolved detailed wave patterns in the eastern part of the Gulf of Finland.
\item The significant wave height is more accurately predicted than the peak period.
\item A major part (84\%) of the wave events with a significant wave height over 7 m occurred between November and January during 1965--2005. None took place before September or after April.
\item The longer fetch components along the axis of the Gulf of Finland displayed an insufficient amount of energy during the highest sea states, which lead to an underestimation of the highest peak periods.
\item Switching off refraction or refining the spatial bathymetric resolution improves the prediction of peak period in the Gulf of Finland. As adding a limiter to refraction is an (unwanted) artificial solution, further investigations are recommended into the physical process of (linear) refraction and physical processes affecting the amount of low-frequency wave energy vulnerable to refraction. 
\item The measurements gaps before and after the ice time can have a significant impact on extreme wave statistics, but the influence on the mean statistics in the hindcast was small. The approach of filling in the gaps with a reliable model hindcast should be considered as a possible solution.
\end{itemize}
On the whole, the hindcast gave an accurate description of the wave climate in the Baltic Sea in offshore areas. The validation against data from near-shore stations also showed an improvement compared over the findings presented in a previous study, which we mainly attribute to a higher spatial resolution in the wave model. Further research into the role of the bathymetry and the suitable value of the refraction limiter is recommended to determine the exact cause of the underestimated long-wave energy in the narrow Gulf of Finland. 

\section*{Acknowledgements}
We acknowledge the work done by Mr. Hannu Jokinen (FMI) in processing the wave buoy data. The allocation of computing time from the High Performance Computing cluster at the Weather Service of the Estonian Environment Agency is also gratefully acknowledged. We would also like to extend our thanks to SMHI, the University of Tartu and the University of Miami (RSMAS) for providing the ice data, wind data and ASIS wave buoy data. Lastly, we want to thank the anonymous reviewers. Their suggestions and comments helped us clarify and strengthen this article. This work was partially funded by Arvid och Greta Olins fond (Svenska kulturfonden, 15/0334-1505) and by CMEMS COPERNICUS Grant WAVE2NEMO.

\section*{References}

\bibliography{Bjorkqvist}
\section*{Appendix}
The effect of the depth-induced refraction on the modelled spectra, significant wave height and peak period was studied with two additional model runs for the 2001 storm in the Gulf of Finland. One model run was made without the depth-induced refraction and the second was made with a high-resolution 500 m bathymetrical grid from the Baltic Sea Bathymetry Database \citep{BSBD}.

\begin{figure}
\begin{center}
\includegraphics[width=14cm]{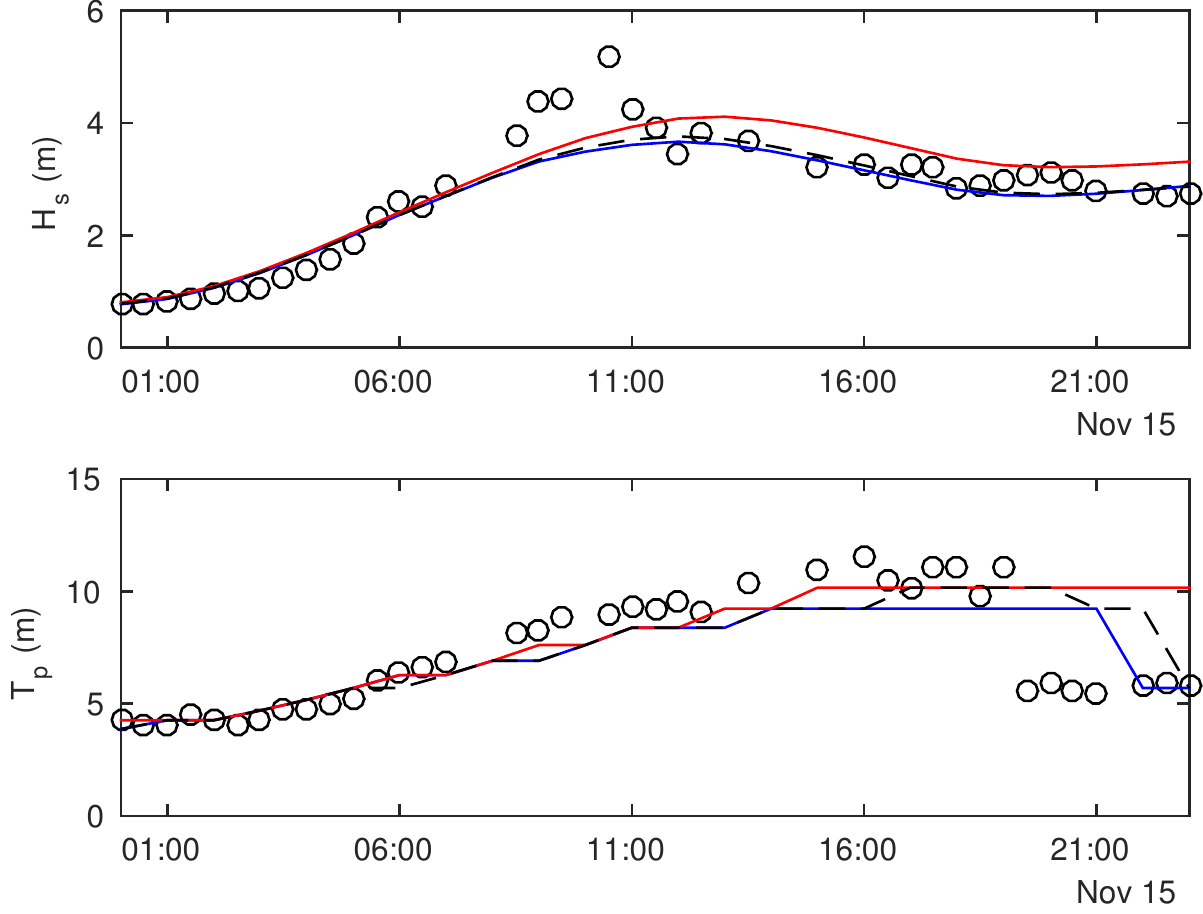}
\caption{The significant wave height (top) and peak period (bottom) at the GoF wave buoy during the November storm in 15 November 2001. Wave buoy (black circles), SWAN (blue), SWAN without depth-induced refraction (red) and SWAN with a 500 m high-resolution grid (dashed black).}\label{Fig:app_timeseries}
\end{center}
\end{figure}

Turning off the depth-induced refraction increases the significant wave height (Fig. \ref{Fig:app_timeseries}). The highest values of the peak period are also modelled better. For the higher resolution grid there is a negligible impact on the significant wave height. Nevertheless, the maximum values of the peak period is more accurately represented by the model.

\begin{figure}
\begin{center}
\includegraphics[width=14cm]{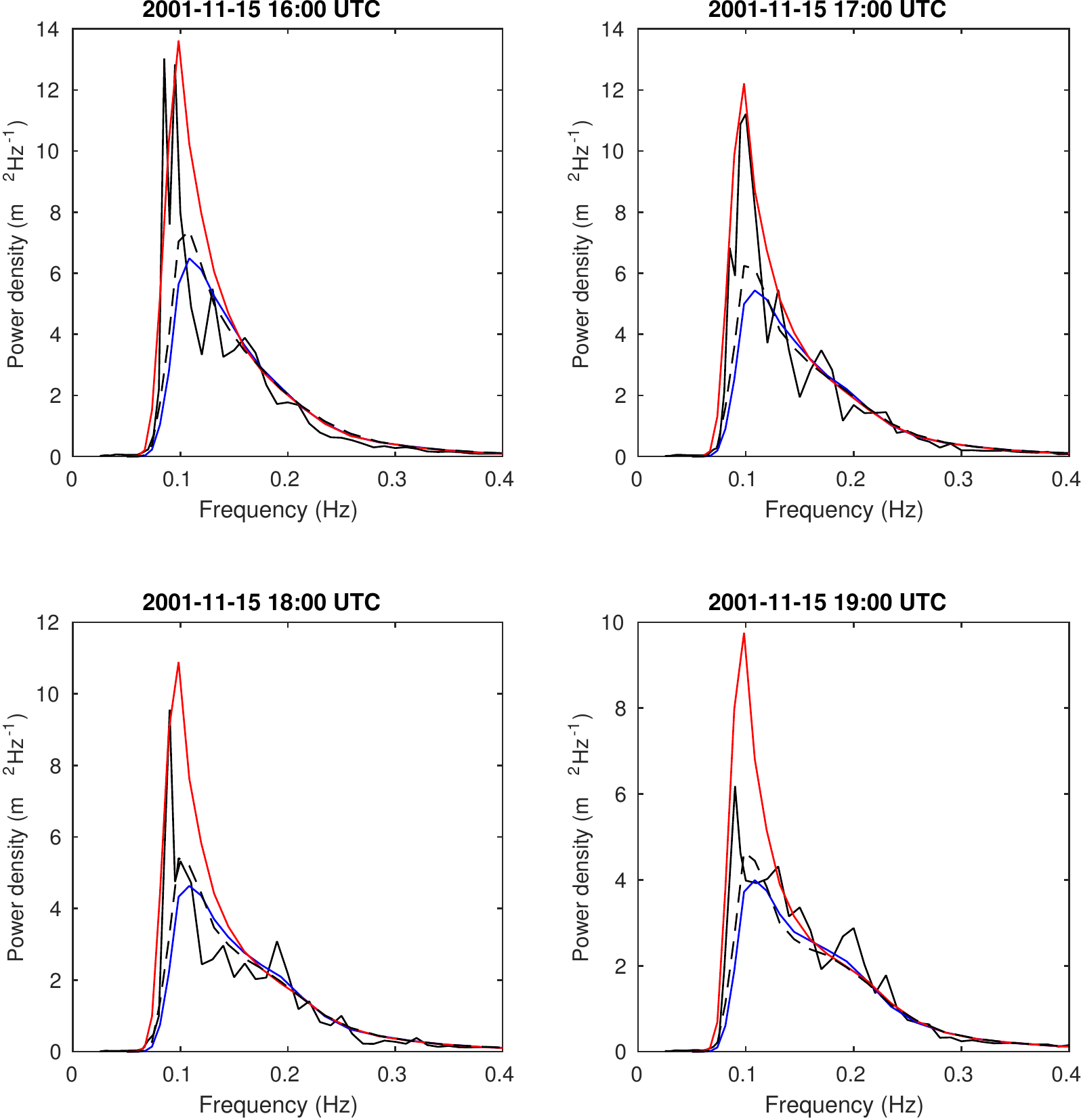}
\caption{Wave spectra at the GoF wave buoy during the November storm in 15 November 2001. Wave buoy (black), SWAN (blue), SWAN without depth-induced refraction (red) and SWAN with a 500 m high-resolution grid (dashed black).}\label{Fig:app_4spec}
\end{center}
\end{figure}

The wave spectra illustrates the difference in performance at this location more clearly (Fig. \ref{Fig:app_4spec}). It is evident that the energy at the lowest frequencies are not modelled correctly with the regular SWAN setup, but the energy in this frequency range is captured when the model is run without the refraction term. While the overall impact when using the higher resolution bathymetrical grid is small, the increase in energy is concentrated to the lowest frequencies. Since all other settings are unchanged, this behaviour indicates that the lacking low-frequency energy in the standard SWAN setup is partly caused by the depth-induced refraction in the model turning away lower-frequency energy from the measurement location.

\end{document}